\documentclass{article}

\usepackage{PRIMEarxiv}

\usepackage[utf8]{inputenc} 
\usepackage[T1]{fontenc}    
\usepackage{hyperref}       
\usepackage{url}            
\usepackage{booktabs}       
\usepackage{amsfonts}       
\usepackage{nicefrac}       
\usepackage{microtype}      
\usepackage{lipsum}
\usepackage{fancyhdr}       
\usepackage{graphicx}       
\graphicspath{{media/}}     
\usepackage{amsmath}
\title{InceptNet: Precise and Early Disease Detection Application for Medical Images Analyses}
  
\author{
  Amirhossein Sajedi \\
  Department of Electrical and Computer Engineering \\
  Semnan University \\
  Semnan, Iran\\
  \texttt{amirh.sajedi@semnan.ac.ir} \\
  \And
  Mohammad Javad Fadaeieslam \\
  Department of Electrical and Computer Engineering \\
  Semnan University \\
  Semnan, Iran\\
  \texttt{fadaei@semnan.ac.ir} \\
}

\begin{document}
\maketitle

\begin{abstract}

In view of the recent paradigm shift in deep AI-based image processing methods, medical image processing has advanced considerably. In this study, we propose a novel deep neural network (DNN), entitled InceptNet, in the scope of medical image processing, for early disease detection and segmentation of medical images in order to enhance precision and performance. We also investigate the interaction of users with the InceptNet application to present a comprehensive application including the background processes, and foreground interactions with users.
Fast InceptNet is shaped by the prominent U-net architecture, and it seizes the power of an Inception module to be fast and cost-effective while aiming to approximate an optimal local sparse structure. Adding Inception modules with various parallel kernel sizes can improve the network’s ability to capture the variations in the scaled regions of interest. To experiment, the model is tested on four benchmark datasets, including retina blood vessel segmentation, lung nodule segmentation, skin lesion segmentation, and breast cancer cell detection. The improvement was more significant on images with small-scale structures. The proposed method improved the accuracy from \textbf{0.9531}, \textbf{0.8900}, \textbf{0.9872}, and \textbf{0.9881} to \textbf{0.9555}, \textbf{0.9510}, \textbf{0.9945}, and \textbf{0.9945} on the mentioned datasets, respectively, which show outperforming of the proposed method over the previous works. \par 

Furthermore, by exploring the procedure from start to end, individuals who have utilized a trial edition of InceptNet, in the form of a complete application, are presented with thirteen multiple-choice questions in order to assess the proposed method. The outcomes are evaluated through the means of Human-Computer Interaction. Considering the distribution of responses in this questionnaire, intriguing findings have been extracted which will urge the future version of this application to advance in the direction of Users' demands documented in this paper.\par

\end{abstract}


\section{Introduction}
Early detection of the diseases and identifying the region of interest play an important role in reducing disease progression and the treatment of patients. The increasing capacity of computer hardware and advancements in algorithms make it possible to design innovative tools for medical image interpretation, for instance, disease diagnosis, and prevent costly medical tests such as Incisional Biopsy. The important part in the diagnosis and treatment of disease through analyzing Medical images exists in precise segmentation and interpretation of regions of interest. Consequently, Computer-Aided Diagnosis (CAD) analysis of medical images can significantly reduce the time, cost, and human-based errors associated with processing the images \cite{lee2019machine}. In this regard, CAD is utilized with two main goals: segmentation and disease diagnosis. The objective of segmentation is to classify an input image into some distinct regions, which puts the pixels into some clusters \cite{mcguinness2010comparative, Amir2019}. In medical applications, the goal is to determine some regions of interest, such as tumor cells \cite{codella2018skin}, body parts \cite{ronneberger2015u}, etc. The interpretation of medical images has a significant effect on the diagnosis and treatment of diseases \cite{lee2019machine, zhao20183d}. Manual interpretation of medical images requires extraction of hidden information in the images which is a tiresome task for humans and involves human errors. Deep learning algorithms, such as Fully Convolutional Neural Networks (FCN)\cite{long2015fully} and Convolutional Neural Networks (CNN), can automatically extract features to analyze the inputs. These algorithms are extensively used for body part segmentation \cite{lee2019machine, azad2019bi, roth2015anatomy}, disease stage classification \cite{yonekura2018automatic}, disease invasiveness\cite{zhao20183d}, and cancer detection \cite{dabeer2019cancer} in biomedical fields. These methods have shown promising results and achieved better results compared to humans. \par

An important problem in medical image interpretation, after classifying it as suspicious or clean, is to locate the disease pattern in the image, if exists. CNN is one the most used deep learning algorithms for classification, where a single label is assigned to the input image. A problem with CNNs is the spatial information loss when the extracted features are fed into fully connected layers. The FCNs are introduced to overcome this deficiency with their pixel-wise classification or segmentation of the images \cite{long2015fully}. Ronneberger et al. \cite{ronneberger2015u} proposed an improvement upon FCN namely U-net for medical image segmentation. One of the advantages of this architecture is its compatibility with a few training examples and its ability to use global location and context information simultaneously. The U-net architecture consists of two paths, including the contraction and expansion paths. In the contraction path, convolutional operations generate feature maps with reduced dimensionality. The expansion path returns the feature maps to the same size as the input image and produces segmentation maps using up-convolutions. There are many extensions of U-net, most of which are related to the concatenation of the feature maps in the skip connections. Azad et al \cite{azad2019bi}. used bi-directional convolutional LSTM in concatenating the feature maps in the skip connections. They also used batch normalization and densely connected convolutional layers in the last convolutional layer of the U-net structure, which helped them achieve better results compared to those obtained by the original U-net architecture. However, the drawback of their proposed architecture was its inability to capture the scale variations in the regions of interest. \par
One of the challenges in medical image segmentation using algorithms is the scale variations in the regions of interest. According to the literature, CNNs, and FCNs used to solve the problem of medical image segmentation have used convolutional layers with fixed kernel sizes at each step of convolution and up-convolution. Using fixed-size kernels at each step cannot deal with the problem of scale variations in the regions of interest.  In this study, Inception modules are incorporated with U-net entitled InceptNet to tackle this problem. Inception modules not only solve the problem of scale variations but also can go deeper while preserving computational efficiency which is a common challenge in the training and prediction phases of deep learning algorithms. The proposed architecture also takes benefit from bi-directional convolutional LSTM for feature concatenation in the skip connections, densely connected convolutions, and batch normalization as proposed in \cite{azad2019bi}. The proposed model is evaluated on four datasets: Breast Cancer Cell segmentation, retina blood vessel segmentation, skin lesion segmentation, and lung nodule segmentation. \par

Scientific Modeling of the proposed method and the interaction of users to this method is another important part that we investigated in this paper. In current Human-Computer interaction (HCI) research, Questionnaires are a powerful tool for measuring the interaction between the application and users to assess user perceptions, preferences, and needs \cite{lazar2017research}. As AI applications become more prevalent, gaining user feedback early in the design process is critical to creating technologies that are trustworthy, ethical, and valuable to society. \par

\begin{figure}[b]
\centering
\includegraphics[width=15.5cm,height=5cm]{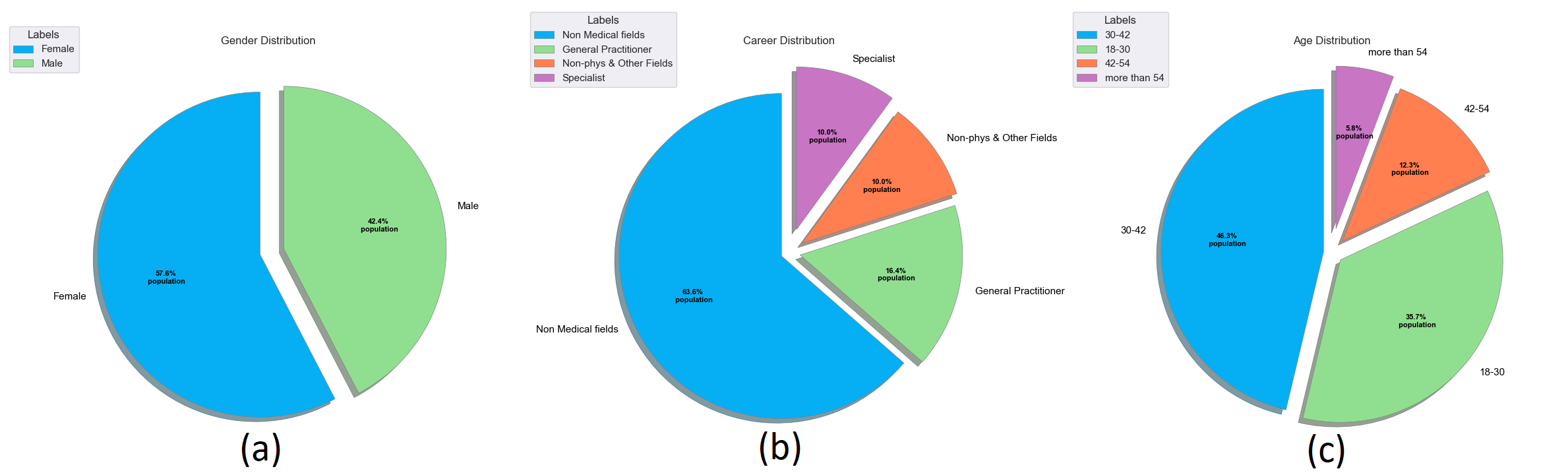}
\caption{a) The Gender distribution, the gender of responders based on their identification, b) Career distribution, consisting of four categories ('non-Medical fields', 'General Practitioner', 'Non-Physician and Other Fields Experts' [e.x. engineer master's graduate], 'Specialist'), c) Category of Age groups of responders including four categories ('18-30', '30-42', '42-54', 'more than 54').}
         \label{Figure_0}
\end{figure}

The questionnaire we designed draws on best practices from HCI literature, using close-ended Likert scale questions to quantify opinions on the proposed InceptNet across several dimensions like interface, trust, and willingness to use \cite{lewis1993defining}. The methodology aligns with similar studies evaluating user acceptance of emerging technologies like smart assistants \cite{lopatovska2019talk}. The data was collected through an online questionnaire filled in by participants at National Libraries, universities, and social media groups. The collected dataset contains 538 responses across 14 attributes, including gender, age, career, and individual opinions on nine aspects of Incept-Net. The opinion questions use a 5-point Likert scale from "lowest" to "highest" in terms such as Agree, Likely, and Low or High. The characteristic features of the Dataset, including a) Gender, b) Age, and c) Career, are illustrated in Figure \ref{Figure_0} in the form of Pie Charts. \par

In short, in this paper, our contribution is summarized as follows.
\begin{itemize}
    \item Presenting a new DL architecture using a particular Inception module, including scales with multiple kernels at the same level of the network.
    \item Employing linear transformation by the Dense layers in skip connections to boost information flows through the layers in our proposed method.
    \item Explaining and Comparing the performance of the proposed method with the existing methods, especially in detecting Cancer Cells.
    \item Examining the interaction between (different types of) users and our launched alpha-version application in workplace environments through HCI perspectives.
    \item Analyzing users' preferences and demands and discussing future prospects.
\end{itemize}

\section{Related works}
    Considering the complexity and time needed for the manual segmentation of medical images, large numbers of automatic methods have been developed to perform this task. In the early stages, some semi-automated and automated methods based on simple rules, such as region growing approaches \cite{haralick1985image}, threshold approaches \cite{sahoo1988survey}, etc., were used, which could not show robust performances when tested on various datasets \cite{pham2000current}. Following the methods above, algorithms were more adaptive when proposed. Jabarouti Moghaddam and Soltani-Zadeh \cite{moghaddam2009automatic} used a two-stage Multi-Layer Perception Artificial Neural Network (MLP-ANN) for brain structure segmentation. Their method consisted of two MLP-ANNs and the output of the first stage was used as the input of the second stage. However, their proposed architecture required manual feature extraction from the input to be fed into the network. Mesejo et al \cite{mesejo2015biomedical}. proposed a geometric deformable model able to combine region-and-edge-based information relying on prior shape knowledge. They used metaheuristic algorithms to tune the model parameters. However, these types of methods were vulnerable to human biases and were unable to handle the variances in real-world data. In recent years, deep learning algorithms have been extensively used for medical image segmentation with outstanding performances compared to human and early-stage algorithms. Convolutional Neural Networks (CNNs) \cite{lecun1998gradient} has been one of the most successfully used deep learning algorithms in computer vision problems. CNNs have been used in many tasks, such as object classification both in images \cite{zhao2019object, krizhevsky2017imagenet} and videos \cite{jang2015object}, object localization \cite{sermanet2013overfeat}, etc. CNNs have dramatically influenced the task of semantic segmentation of medical images in recent years. Ciresan et al \cite{ciresan2012deep}. utilized a CNN with sliding windows for semantic segmentation of the neuronal structure of the pixel in the center of the sliding window. In an automatic segmentation of brain MRI images, Cui et al \cite{cui2016brain}. employed a CNN and extracted some patches from the images for training the model. A probabilistic bottom-up approach was proposed by Roth et al \cite{roth2015anatomy}. for the segmentation of the pancreas in abdominal CT images. Müller and Kramer \cite{muller2021miscnn} introduced a framework, named Medical Image Segmentation Convolutional Neural Network (MIScnn) that provided a pipeline for data I/O, preprocessing, data augmentation, patch-wise analysis, and metrics for segmentation of medical images. However, these networks required large datasets to be trained and suffered from loss of spatial information when the extracted features were fed into the fully connected layers.\par
    A frequently used neural network for medical image segmentation tasks is Fully Convolutional Networks \cite{isensee2018nnu, zhou2017fixed}. Zhou et al \cite{zhou2016three}. segmented the anatomical structures on 3D Computed Tomography images using an FCN. The model was trained end-to-end to perform voxel-wise multiple-class classification which assigned an anatomical label for each voxel in an input image. A very deep FCN was proposed by Drozdzal et al \cite{drozdzal2016importance}. which utilized short skip connections. They showed that the proposed model with both short and long skip connections outperforms the original model with only short skip connections. An improved version of FCN, namely SegNet was proposed by Badrinarayanan et al. \cite{badrinarayanan2017segnet}. The proposed architecture was composed of 13 encoder layers for extraction of spatial features and corresponding 13 decoder layers to predict the segmentation masks. Bai et al \cite{bai2018recurrent}. proposed a combination of FCN and Recurrent Neural Network (RNN) for the segmentation of medical image sequences. Their network was able to consider both spatial and time-dependent information in MR images. Lei et al \cite{lei2021defed}. improved the accuracy of brain tumor segmentation and liver tumor segmentation with the aid of the superposition of multi-scale atrous convolutions. Du et al \cite{du2022net}. introduced the asymmetric lightweight medical image segmentation network (AL-Net) to speed up the inference process of segmentation. Their proposed network was tested on three datasets, including the retinal vessel, cell contour, and skin lesion segmentation, and showed its superiority to previous architectures in terms of accuracy and speed. However, their network was unable to capture variations in the scale of the regions of interest. Tang et al \cite{tang2022unified}. proposed a novel end-to-end architecture, namely Uncertainty Guided Network (UG-Net), to solve the problem of boundary information loss due to consecutive pooling layers and convolution strides that result in unreliable boundaries in segmentation. Their method consisted of three parts, including the coarse segmentation module (CSM) that generated the coarse segmentation and the uncertainty map, an uncertainty-guided module (UGM) that was used to leverage the uncertainty map in an end-to-end manner, and a feature refinement module embedded with several dual attention (DAT) blocks to generate the final segmentation. Results showed that the proposed method was superior to other state-of-the-art networks on nasopharyngeal carcinoma (NPC) segmentation, lung segmentation, optic disc segmentation, and retinal vessel segmentation.\par
    
Despite the outstanding performance of CNN and FCN architecture in medical image segmentation tasks, their major drawback was the unavailability of large datasets to train the model. U-net is an FCN architecture proposed by Ronneberger et al. \cite{ronneberger2015u} to mitigate this drawback. U-net is similar to FCN and SegNet \cite{badrinarayanan2017segnet} that consists of entirely convolutional layers with encoding and decoding paths used for semantic segmentation tasks. The encoding path is responsible for extracting feature maps with reduced dimension and the decoding path produces segmentation maps using up-convolutions. An advantage of U-net over other FCNs is that it is compatible with a few numbers of training data. Over the years, many extensions of U-net have been proposed. An extension of U-net was proposed by Milletari et al \cite{milletari2016v}. named V-net for 3D image segmentation of MRI volumes. Azad et al \cite{azad2019bi}. proposed BCDU-net as an extension to U-net. This architecture used Bi-directional convolutional LSTM (BConvLSTM) in the skip connection and densely connected convolutional blocks in the last layer of the encoding path with the purpose of feature propagation and feature reuse. Since the feature maps from the contraction path have higher resolution and the feature maps extracted from the corresponding expansion path have more semantic information, BConvLSTM helps combine these feature maps with better segmentation results by providing a non-linear way to concatenate the feature maps in the skip connections. This can alleviate the semantic gap between the feature maps from the early layers of the encoder path and the deeper layers of the decoder path. They also believed that densely connected convolutions improve information flow through the network and mitigate learning redundant features \cite{maleki2023}. However, they used fixed-size convolutional layers in the convolutional blocks which could not capture the variations in the scale of the regions of interest. In another work, Ibtehaz and Rahman \cite{ibtehaz2020multiresunet} utilized a chain of convolutional layers with residual connections to concatenate the feature maps in the skip connections and called it ResPath. Some other utilizations of U-net in medical applications are segmentation of biomedical images of liver \cite{christ2016automatic}, skin lesion \cite{lin2017skin}, kidney \cite{cciccek20163d}, lung nodule \cite{setio2017validation}, prostate \cite{yu2017volumetric}, etc.\par
    In this paper, an extension of the U-net called IncepNet is proposed for further improvement of the efficiency and performance of the network. In this architecture, Inception modules replace the sequence of two convolutional layers in the original U-net. Also, as proposed in BCDU-net \cite{azad2019bi}, BConvLSTM is used for concatenation in skip connections, dense connections in the last layer of the encoding path, and batch normalization in the decoding path. However, their proposed architecture was unable to capture the variations in the regions of interest. In this study, the Inception module is added to the structure of the network to cope with this limitation. The proposed architecture is tested using four tasks of segmentation, including the segmentation of retinal blood vessels, skin lesions, lung nodules, and breast cancer cells. Since some functionalities of the BCDU-net \cite{azad2019bi} and the original U-net \cite{ronneberger2015u} are used in the proposed architecture; their structure is briefly described in the following. More details can be found in \cite{azad2019bi}.\par



\section{Baselines Examination}
    As mentioned earlier, BCDU-net is an extension of U-net which has a similar architecture to FCN and SegNet. BCDU-net has a symmetric architecture with encoder and decoder paths as the original U-net. These parts are briefly described in the following subsections. \par

\begin{figure}[tb]
\begin{center}
\includegraphics[width=15cm]{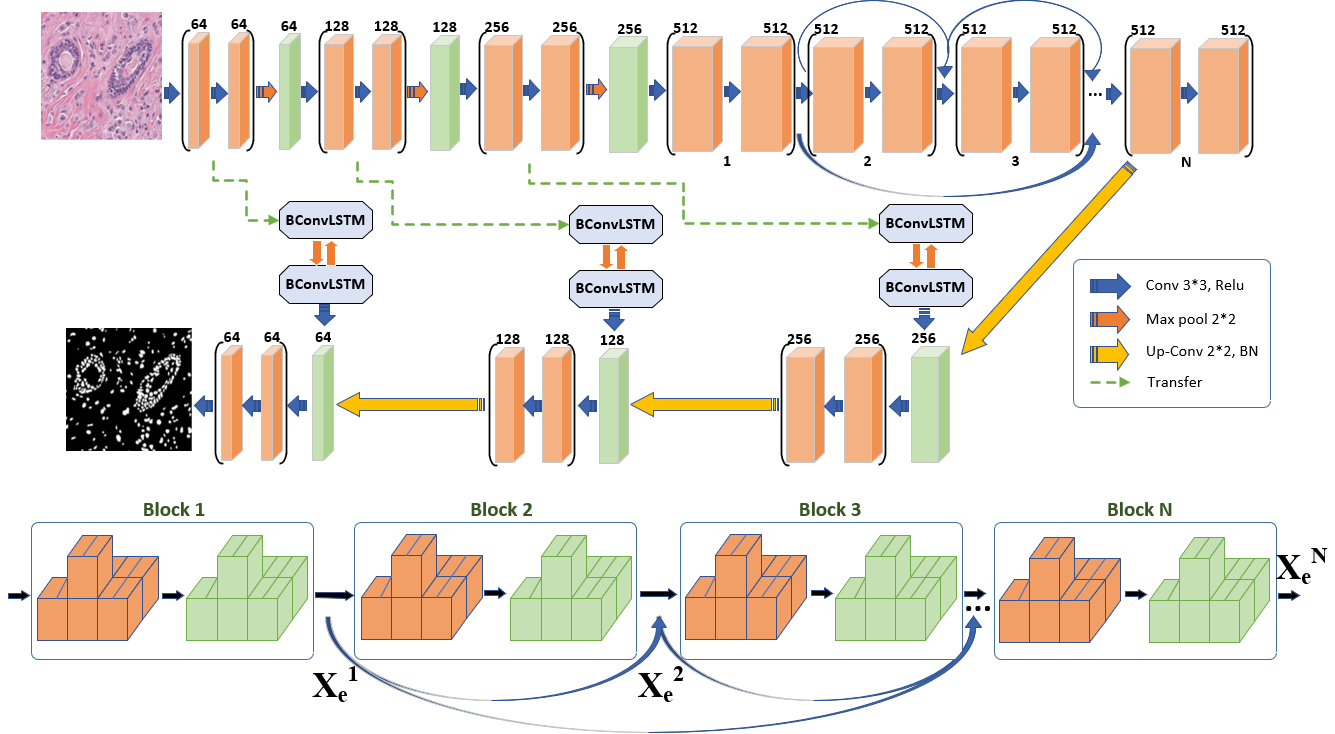}
\end{center}
\caption{Base structure of the InceptNet before adding Inceptions and skip connections that are inspired by BCDU-net\cite{azad2019bi}.}
\label{Figure_1}
\end{figure}

In the baseline architecture (Figure \ref{Figure_1}), each cube includes a multi-channel feature map. The number of channels and axis size are defined on top of the cube. This architecture comprises continuous layers of a 3×3 convolutional network followed by the ReLU activation function and a down-sampling function, which uses a 2×2 max pooling operation with 2 strides. At every sampling step, the number of feature channels is doubled. On the other hand, the expansive path consists of feature channels for back-sampling with a 2X2 convolution that reduces the feature channel to halves. Also a series with the corresponding path cut feature map from the contracting path and two 3X3 convolutions, each followed by a ReLU activation function and the cutting is necessary because of border pixel losses at every convolutional step. At the end layer, the 1×1 convolution is used to map each component feature vector to its desired number of classes, adding up to 23 convolutional networks in a U-net model \cite{ronneberger2015u}. \par
    
BCDU-net is inspired by U-net with densely connected convolutional networks, with increased performance for segmenting medical images. The dense Layer of the BCDU-net consists of four steps; every step has two convolutional networks with 3x3 filters along with a 2x2 max pooling layer with activation function ReLU. At the end of each step, the feature map gets doubled \cite{azad2019bi}.\par

\subsection{Encoder}
    The encoder part of the network is a stack of convolutional blocks with 3×3 convolutional layers. After each convolutional block, a max-pooling layer with a pooling size of 2×2 and a stride of 2 is placed. This is repeated four times, and the number of feature maps after each block is doubled. In the BCDU-net, after the last block in the down-sampling section, a sequence of densely connected layers is placed. In this part of the network, the output of the previous convolutional block is concatenated with the output of the current convolutional block and passed to the next block as its input. The number of filters in this part remains constant. The advantage of using densely connected convolutions is that it can avoid learning redundant features and also it allows information flow through the network and feature reuse. Moreover, densely connected convolutions can use all previously produced feature maps, which can alleviate the risk of vanishing or exploding gradients during learning.\par

\subsection{Decoder}
    In the decoding path, the feature map coming from the previous step is up-sampled using a transposed convolution with a size of 2×2 that reduces the number of feature maps by half. Like the encoding path, the output of the up-sampling operation is passed through a block of two 3×3 convolutional layers and it is repeated four times. In the end, a 1×1 convolution assures that the dimension of the final segmentation output resembles the input size. It should be pointed out that the activation function for all convolutional layers, except for the final layer, is ReLU, and for the final layer is sigmoid. As in the original U-net architecture, the extracted feature maps from the encoding and decoding path are concatenated in skip connections. However, instead of simple concatenation, BCDU-net uses a BConvLSTM to concatenate these feature maps. \par
    One more point about the BCDU-net is Batch Normalization (BN) \cite{ioffe2015batch} in the decoding path. In the decoding path, after each up-sampling, the output is gone through a batch normalization process. Batch normalization enhances the network’s stability and increases the training speed. Figure \ref{Figure_1} shows the structure of the BCDU-net. The difference between the BCDU-net and the original U-net is that the U-net does not use BConvLSTM for concatenation and dense connections. Also, batch normalization is not utilized in the U-net. In  Figure \ref{Figure_1}, the network architecture is shown in the upper section, and the lower section shows the densely connected convolutions. More details can be found in \cite{azad2019bi}.\par

\subsection{Bi-Directional ConvLSTM}
    As mentioned earlier, BCDU-net uses Bi-Directional ConvLSTM as the skip connections for feature map concatenation. In this regard, the output of the batch normalization step goes through a BConvLSTM layer.  Convolutional LSTM has the advantage of taking into account the spatial correlations which the standard LSTM does not. ConvLSTM is composed of an input gate, \(i_{t}\), an output gate, \(o_{t}\), a forget gate, \(f_{t}\), and a memory cell, \(C_{t}\). The mathematical formulation for the ConvLSTM is given as follows \cite{azad2019bi}: \par

\vspace{0.3cm}
    \[ i_{t}=\sigma (W_{xi}* X_{t}+W_{hi}*H_{t-1}+W_{ci}*C_{t-1}+b_{i} )  \]
    \[ f_{t}=\sigma (W_{xf}* X_{t}+W_{hf}*H_{t-1}+W_{cf}*C_{t-1}+b_{f} )  \] 
    \[C_{t}=f_{t} \odot C_{t-1}+i_{t} tanh(W_{xc}*X_{t}+W_{hc}*H_{t-1}+b_{c}) \]
    \[ o_{t}=\sigma (W_{xo}* X_{t}+W_{ho}*H_{t-1}+W_{co} \odot C_{t}+b_{c} ) \] 
    \[ H_{t}=o_{t} \odot tanh_{(C_{t})} \]  
    
\vspace{0.3cm}

    In the above equations, * and \(\odot\) are the convolution and Hadamard functions, respectively. t is the current time step \(t\in{1,…, N}\) at which N is the total number of time steps, \(H _{t}\)is the hidden state tensor, \(X _{t}\)denotes the input tensor, and \(W _{x*}\),\(W _{h*}\) and \(W _{c*}\) are two-dimensional convolution kernels of the input, hidden and memory cell state, respectively, and \(b _{i}\),\(b _{f}\),\(b _{o}\) and \(b _{c}\) are the bias terms. All \(W _{**}\) and \(b _{*}\) parameters are learned during training the network. Finally, \( \sigma\) denotes the sigmoid activation function.\par
    In BConvLSTM, two ConvLSTMs are used for processing the data in forward and backward directions. This can take into account the dependencies in both directions. It should be noted that ConvLSTM considers only the forward dependencies in the input data. However, backward dependencies are needed to consider all the information in a data sequence. Therefore, two ConvLSTMs are used for the forward and backward processing of the inputs. As a result, there will be two sets of parameters each of them for forward and backward states. In this manner, the output of the BConvLSTM is given as below:\par

    \[Y _{t}=tanh (W _{y}^{\overrightarrow{H}} * \overrightarrow{H} _{t}+W \_{y}^{\overleftarrow{H}}{\overleftarrow{H} _{t}}+b) \]

\begin{figure} [tb]
\begin{center}
\includegraphics[width=12cm]{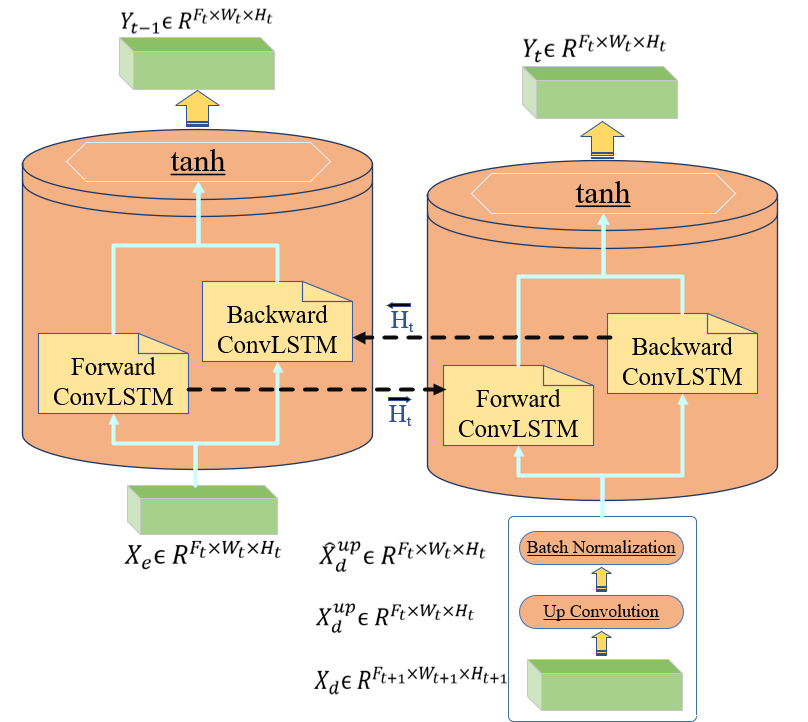}
\end{center}
\caption{A scheme of Bi-Directional ConvLSTM, and the information flow between two layer.\cite{azad2019bi}.}
\label{Figure_2}
\end{figure}

    Where \(Y _{t} \in R^{F _{l}*W _{l}*H _{l}}\) is the final output of the BConvLSTM at time step t, \( \overrightarrow{H} _{t}\) and \(\overleftarrow{H} _{t}\) are the forward and backward hidden state tensors at time step t, respectively and b denotes the bias term. \(W _{y}^{\overrightarrow{H}}\) and \(W _{y}^{\overleftarrow{H}} \) are two-dimensional convolution kernels of the forward and backward hidden state tensors, respectively. The structure of the BConvLSTM is depicted in Figure \ref{Figure_2}. Interested readers can refer to \cite{ronneberger2015u} and \cite{azad2019bi} for more details on the U-net and BCDU-net architectures.\par

\par
\section{Proposed Method}

\subsection{Inception modules}
    Commonly, the regions of interest in the segmentation of medical images have irregular shapes and varying scales. Consequently, these variations have to be taken into account to make the predictive network more robust. Inception modules, introduced by Szegedy et al \cite{szegedy2015going}. can inspect the objects of interest with different scales by parallelizing convolutional layers with kernels of different sizes. The output of these layers is then merged and sent to deeper layers. This characteristic of the inception modules allows the network to go deeper with not much increase in computational expense.\par
\begin{figure} [tb]
\begin{center}
\includegraphics[width=11cm]{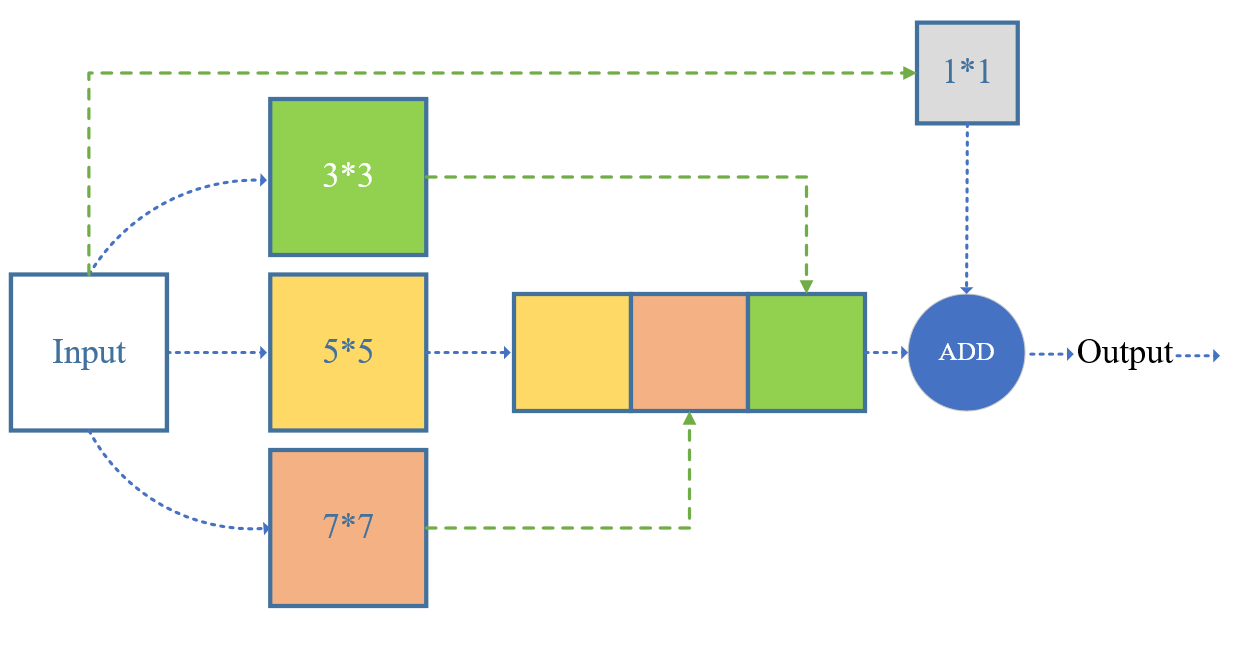}
\end{center}
\caption{A regular inception block.  \cite{szegedy2015going}}
\label{Figure_3}
\end{figure}
\begin{figure} [tb]
\begin{center}
\includegraphics[width=11cm]{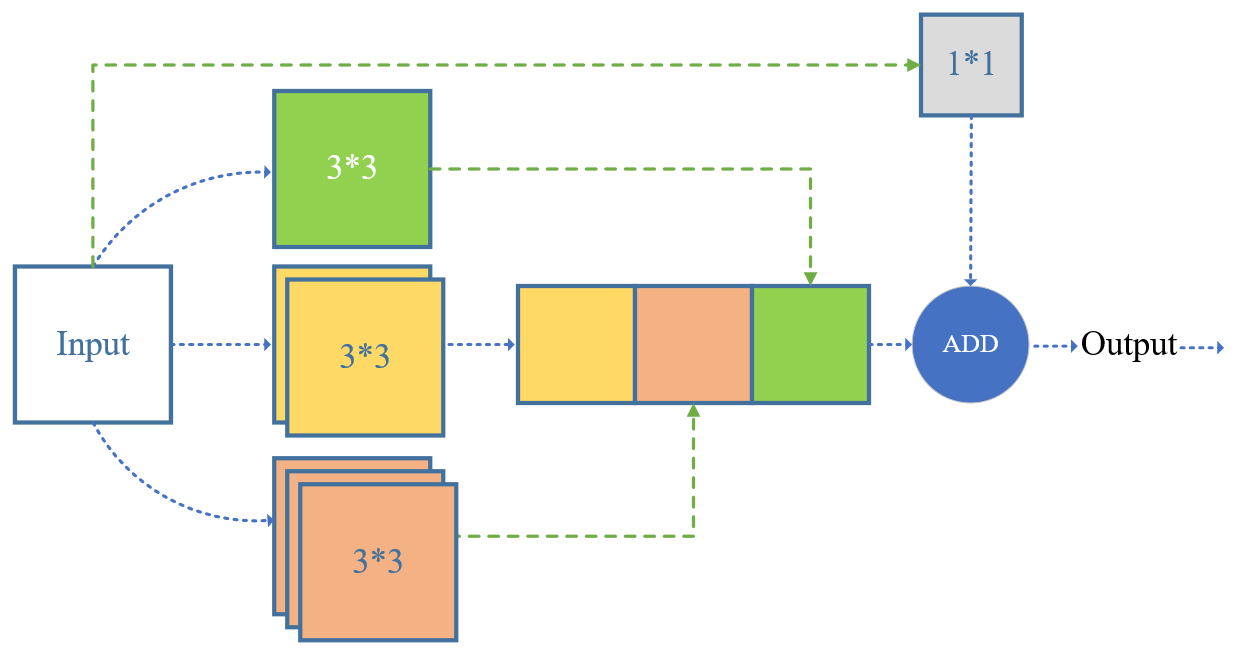}
\end{center}
\caption{Inception block with efficient stacked convolutional layers, the novel proposition of our proposed method.  \cite{szegedy2015going}.}
\label{Figure_4}
\end{figure}

    Recalling the structure of the U-net, there was a block of two 3×3 convolutional layers in the encoding and decoding paths. This sequence of two 3×3 convolutions resembles a 5×5 convolution, and subsequently, a sequence of three 3×3 layers resembles a 7×7 convolutional layer. Therefore, a reasonable way to implement a U-net architecture combined with Inception modules is to consider a block of 3×3 and 7×7 layers in parallel to a 5×5 convolutional layer. It is common to add a residual 1×1 convolution because it may result in better extraction of spatial information. As a result, putting the Inception blocks instead of a series of convolutional layers in the structure of the network enables the network to handle the regions of interest of different scales (Figure \ref{Figure_3}). One important point to take into account is that by adding extra convolutional layers in parallel, there will be an increase in the required memory (this limitation is more problematic on personal computers). For example, the proposed architecture with Inception blocks, as shown in Figure \ref{Figure_3}, has approximately 40 million parameters. Consequently, we replace the 5×5  and 7×7 layers, which are more computationally expensive, with a series of 3×3 convolutions (Figure \ref{Figure_4}). Here, the output of the second and the third 3×3 layers is an approximation of the output of the 5×5 and 7×7 convolutions, respectively.\par
    
\subsection{InceptNet}
    In the InceptNet architecture, the sequence of two convolutional layers is replaced with an Inception module as described in the previous section (figure \ref{Figure_4}). Depending on the number of channels in the input image, these blocks can have 2D or 3D convolutional layers in parallel. To have a more robust comparison between the performance of the proposed network and the performance of the BCDU-net, we keep the same number of filters in an Inception block and a sequence of two 3×3 convolutional layers in the BCDU-net structure. Moreover, it is tried to keep the number of parameters in the proposed method and the original BCDU-net as close as possible to each other.\par
\begin{figure} [tb]
\begin{center}
\includegraphics[width=17cm]{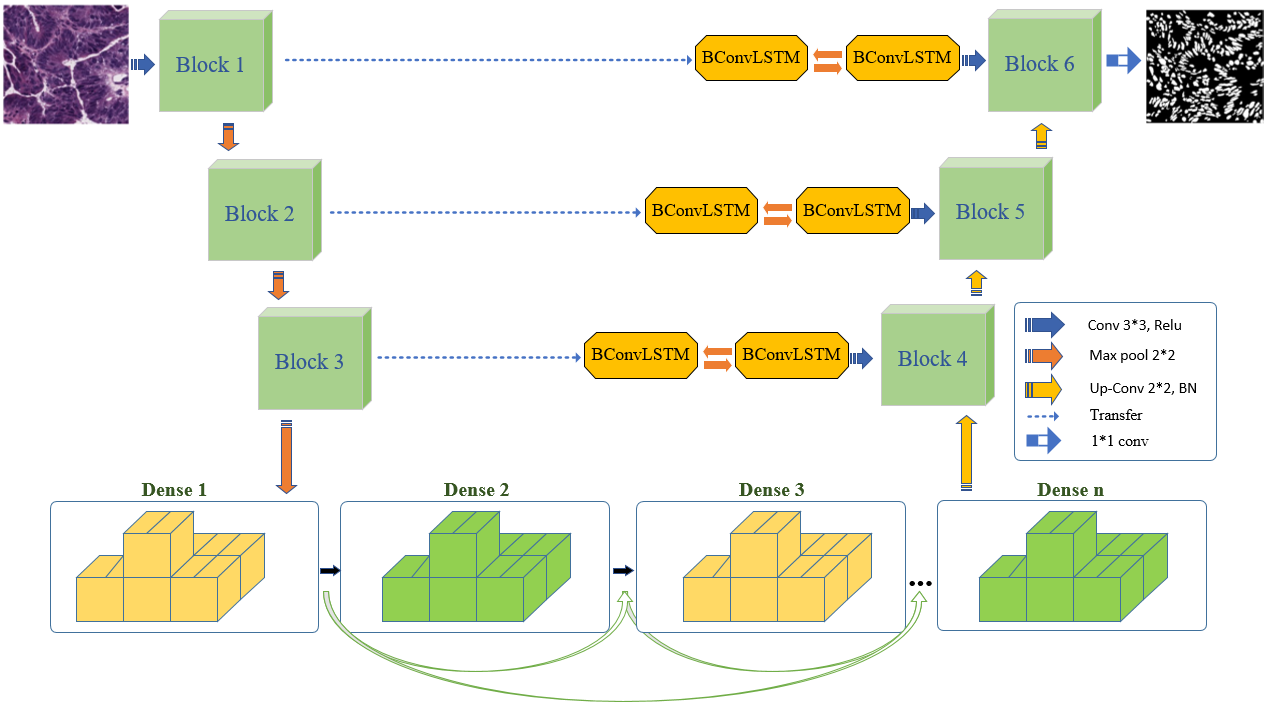}
\end{center}
\caption{InceptNet incorporated with bi-directional ConvLSTM in the skip connections and dense connections.}
\label{Figure_5}
\end{figure}

    Like in the BCDU-net, to mitigate the semantic gap between the feature maps from the encoding and decoding stages in the skip connections, we use BConvLSTM to non-linearly concatenate the feature maps in the skip connections. Also, in the last layer of the encoding path, we use densely connected blocks. In this stage, the output of the previous Inception block is concatenated with the output of the current Inception block, and the outcome is sent to the next block as its input. Like in \cite{azad2019bi}, several 3 densely connected blocks are considered at the last layer of the encoding path. In the proposed method, the activation function at all layers is set to ReLU, except that the activation function of the output layer is set to sigmoid as in the BCDU-net architecture. Other hyperparameters to train the network are chosen as in \cite{azad2019bi}. Figure \ref{Figure_5} shows the schematic of the proposed architecture. In this architecture, Inception blocks (Figure \ref{Figure_4}) are indicated by rectangles (blue for encoding and decoding paths, red for the densely connected convolutions). The dimension of the input and output of the network is the same as the network outputs a labeled image in which the pixels with the same label represent a specific cluster of semantic regions. The inputs are first converted into greyscale except for the ISIC dataset as in \cite{azad2019bi} to make the comparison more robust. The input layer is made using the Input layer of the Keras library to make it compatible with the dimension of the input dataset. To avoid overfitting, the Dropout layer is used in the expansion path of the network. The proposed model is implemented using Python libraries.\par

\section{Datasets}
    The proposed architecture is evaluated on four datasets from multiple modalities: retinal blood vessel segmentation (DRIVE dataset), lung nodule segmentation, skin lesion segmentation (ISIC 2018), and Breast Cancer Cell detection. Each dataset has its own challenges that are implied in the following subsections. In all datasets, 20\% of the training data is used for validation (except for ISIC 2018) and the model with the lowest validation loss is saved as the best model for prediction. All datasets are available and are referenced in the footer. A summary of the utilized dataset is given in Table \ref{tab:table_2}.\par

\subsection{DRIVE dataset}
    The DRIVE dataset\footnote{https://docs.activeloop.ai/datasets/drive-dataset} \cite{staal2004ridge}, consists of 40 color images used for segmentation of retinal blood vessels. This dataset is split into 20 training and 20 testing images. The size of the original images is 565×584 pixels. Note that the input images are first converted to grayscale. A problem with this dataset is the small size of it for training a deep neural network. Consequently, the same strategy is used by Azad et al \cite{azad2019bi}. is used in this study. This strategy is to divide the input images into random patches. Considering a size of 64×64 pixels for each patch, a total number of 200,000 patches are extracted from the training images of which 160,000 patches are used for training and 40,000 patches for validation.\par

\subsection{ISIC 2018 dataset}
    The ISIC dataset\footnote{https://challenge.isic-archive.com/data/\#2018} \cite{codella2018skin} was published by the International Skin Imaging Collaboration (ISIC). This dataset consists of 2594 images of which 1815 images are used for training, 259 for validation, and 520 images for testing along with their ground truth annotations. The images have a dimension of 700×900×3 which is computationally demanding. A challenge with this dataset is the large dimension of the images. The authors in \cite{alom2018recurrent} reduced the size of the images to 256×256×3 pixels using a bi-linear interpolation technique. We applied the same technique to make a better comparison between the results as applied in the original BCDU-net structure. Note that no patches are extracted from this dataset and the original images after re-sizing are used for training, validation, and testing.\par

\subsection{Lung Segmentation dataset}
    The third dataset is a lung nodule segmentation\footnote{https://zenodo.org/record/3723295\#.YtO5rnZByHs} dataset was introduced in the Lung Nodule Analysis (LUNA) competition at the Kaggle Data Science Bowl in 2017. There are 2D and 3D computed tomography (CT) images with their corresponding annotated images for lung segmentation, however, 3D images are used for segmentation in this study. The dataset includes 1021 annotated CT images with a size of 512×512 from which 70\% of the data are used as the train set and the remaining as the test set. As in the DRIVE dataset, the input images are converted to grayscale, but no patches are extracted. The lung region has a Hausdorff value close to other objects like bone and air. Therefore, as in the BCDU-net, the lung region is learned by learning the surrounding tissues following Algorithm 1 in \cite{azad2019bi}. \par

\subsection{Breast Cancer Cell dataset}
    Breast Cancer Cell\footnote{https://www.kaggle.com/datasets/andrewmvd/breast-cancer-cell-segmentation}  consists of 58 H\&E stained histopathology images. This dataset was originally used for the detection of breast cancer cells. From the 58 colored images, 30 images are used for training, and 28 images for testing. The original images have a size of 896×768×3. As in the DRIVE dataset, a grayscale filter is applied to the images, then some random patches of size 128×128 are extracted from them. Totally, 200,000 patches are extracted from which 160,000 are used for training and the remaining 40,000 patches are used for validation.\par

\begin{table}
 \caption{A brief description of the datasets}
  \centering
  \begin{tabular}{llll}
    \toprule
    Dataset            & No. of images  & Original size  & Input size \\
    \midrule
    DRIVE              & 40             & 565×584×3      & 64×64×1    \\
    ISIC 2018          & 2594           & 700×900×3      & 256×256×3  \\
    Lung               & 1021           & 512×512×3      & 512×512×1  \\
    Breast Cancer Cell & 58             & 896×768×3      & 128×128 ×1 \\
    \bottomrule
  \end{tabular}
  \label{tab:table_1}
\end{table}

\section{Experiments}
    In this section, the proposed method is evaluated on four medical datasets described above. Keras with Tensorflow backend is used for the implementation of the model. Other tools used for implementation are listed in Table \ref{tab:table_1}. Some evaluation metrics are used to compare the results obtained from different architectures, including the accuracy (AC), sensitivity or recall (SE), specificity (SP), F1-score, ROC curve, the area under the curve (AUC), and Jaccard similarity (JS). ROC curve is the cross-plot of the true positive rate versus the false positive rate, and AUC refers to the area under the ROC curve that shows the performance of the model at segmenting the inputs. The above metrics are defined in the following.\par   

    \begin{equation}
    AC ={\frac {Total number of true predictions}{Total number of samples}}
    \end{equation}

    \begin{equation}
    SE ={\frac {Number of TP}{Number of TP+number of FN}}
    \end{equation}

    \begin{equation}
    SP ={\frac {Number of TN}{Number of TN+number of FP}}
    \end{equation}

    \begin{equation}
    f1-score =2{\frac {Precision*recall}{Precision+recall}}
    \end{equation}

    \begin{equation}
    Precision ={\frac {Number of TP}{Number of TP+number of FP}}
    \end{equation}

    In the equations above, TP, FP, TN, and FN denote true positives, false positives, true negatives, and false negatives, respectively. If A and B are two matrices, the Jaccard similarity score is defined as the ratio of A and B intersection to A and B union:\par

    \begin{equation}
    JS ={\frac {A\cap B}{A\cup B}}
    \end{equation}

\subsection{Base model}
    Considering the fact that the proposed architecture, IncepNet, is a modification of the state-of-the-art BCDU-net architecture, we compare the performance of the proposed method with the BCDU-net architecture as the base model. The number of filters in the encoder and decoder parts of the model is set to 64, 128, 256, and 512 for the proposed BCDU-net architectures. To have a better comparison, we run the model with (d=3) and without (d=1) densely connected blocks. Table \ref{tab:table_3} shows the number of parameters in both cases to reach the same level of depth. As can be seen, the number of parameters in the proposed model is slightly lower than the number of parameters in the base model (BCDU-net architecture).\par

\begin{table}
 \caption{Number of parameters of the models}
  \centering
  \begin{tabular}{ll}
    \toprule
    Model            & No. of parameters \\
    \midrule
    BCDU-net (d=1)              & 8,205,573 \\
    BCDU-net (d=3)          & 20,659,717 \\
    InceptNet (d=1)               & 7,829,872 \\
    InceptNet (d=3)             & 18,453,190 \\
    \bottomrule
  \end{tabular}
  \label{tab:table_2}
\end{table}

\subsection{Training}
    In fact, semantic segmentation with two classes is a binary classification in which a label is assigned to each pixel being a point of interest or not. As a result, the loss function of the network will be the binary cross-entropy. This function is minimized during the training of the network. The general formulation of the binary cross-entropy is given as follows:\par

    \begin{equation}
    J(Y,\hat Y) = \Sigma _{i \in x} - (y _{i}log(\hat y _{i})+(1- y _{i})log(1- \hat y _{i}))
    \end{equation}

    In the equation above, J is the binary cross-entropy loss function, X is an input image, Y is the corresponding ground truth mask of X, \(\hat y\) is the predicted mask by the predictive network. For pixel i, the predicted label is \(\hat y _{i}\) and the true label is indicated by \(y _{i}.y _{i}\) is binary that takes only 0 or 1, however, \(\hat y _{i}\) is between 0 and 1 which is predicted by the network. Therefore, \(J(Y, \hat Y)\) can be computed with no mathematical problem. If a mini-batch approach with a size of n is used for training, the loss function is calculated over the images inside the batch \((X _{i})\).\par

    \begin{equation}
    J(batch) = 1/n \Sigma _{i=1}^n (Cross Entropy(Y _{i}, \hat Y _{i}))
    \end{equation}

    It is worth pointing out that the model is trained using the Adam optimizer with the hyperparameters as utilized by Azad et al \cite{azad2019bi}. Training stops if the validation loss does not change in 10 consecutive epochs. The maximum number of epochs is set to 50, 100, 50, and 50 for DRIVE, ISIC, Lung, and Breast Cell Cancer datasets, respectively. \par

\section{Results}
    As explained above, the proposed method is tested on four datasets described in section 6. In this section, the performance of the InceptNet and the BCDU-net on the test datasets is compared using the evaluation metrics mentioned in section 7. Moreover, the training and validation accuracy at the end of each epoch is plotted to make sure of the convergence of the model. All the training, validation, and testing strategies are the same as in \cite{azad2019bi} to have a better comparison between the proposed model and the original BCDU-net model. Also, the results on each dataset are compared with the results of other methods in the literature if exist. The results of the models on each dataset are presented in the following.\par

\subsection{Retinal Blood Vessel Segmentation}
\begin{figure}[tb]
    \centering
    \includegraphics[width=13cm, height=3.5cm]{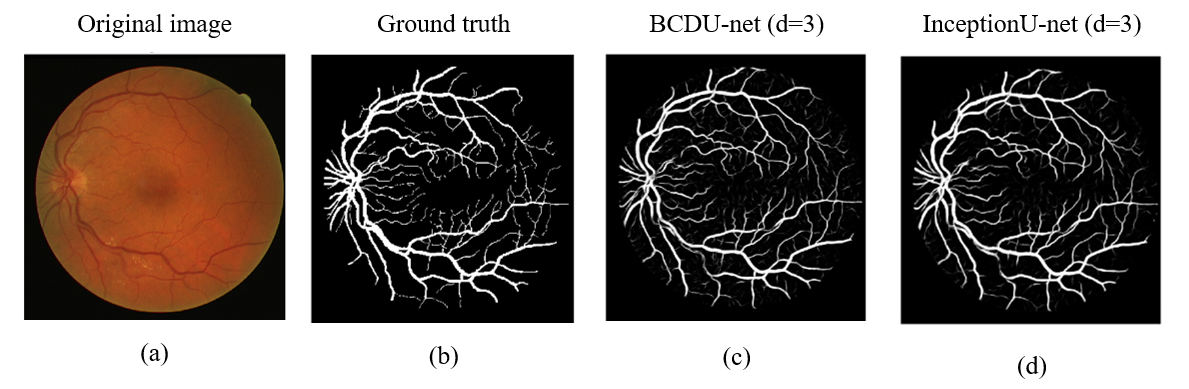}
    \caption{a) original colored image, b) corresponding ground truth mask, c) predicted mask by BCDU-net (d=3), d) predicted mask by InceptNet (d=3) on the DRIVE dataset}
    \label{Figure_6}
\end{figure}

    Figure \ref{Figure_6} illustrates an example of the segmentation results of the best-performing version of the proposed network and BCDU-net (with dense connections) on the DRIVE dataset. The left column in the first row represents the original color image and the right column shows the ground truth mask. The predicted output of both architectures is illustrated in the second row. To have a better comparison between different models and the InceptNet, the final quantitative results are provided in Table \ref{tab:table_4} (qualitative results were not available for other methods). It should be pointed out that the model is trained with d=1 and d=3 as the number of densely connected blocks. In fact, there are no dense connections with d=1 and the architecture is like the original U-net. With d=3 there are three Inception blocks in the last level of the encoder path of which two of them are densely connected blocks. According to the obtained results, the proposed method shows outperformance to the BCDU-net structure for all of the evaluation metrics regardless of the number of dense connections. Additionally, d=3 outperformed the architecture with d = 1. The plot of training and validation accuracy against the epoch number is depicted in Figure \ref{Figure_10}a and Figure \ref{Figure_11}a for d=1 and d=3, respectively, showing that the model converges properly. The reason behind the validation accuracy being larger than the training accuracy in the early epochs is twofold: the small size of the validation data, and the calculation of the validation accuracy at the end of the epoch. In the end, the overall performance of the proposed InceptNet on the DRIVE dataset is shown by the ROC curve in Figures \ref{Figure_12}a and \ref{Figure_13}a.\par

    As shown in Figure \ref{Figure_6}, the output of the original BCDU-net has more tiny structures labeled as retinal blood vessels which is an indication of its inability to deal with small-scale structures.\par

\begin{table}
 \caption{Performance of the different segmentation methods on the DRIVE dataset.}
  \centering
  \begin{tabular}{llllll}
    \toprule
    Method   & F1-score & Sensitivity      & Specificity    & Accuracy       & AUC\\
    \midrule
    COSFIRE filters \cite{azzopardi2015trainable}&    -             &  0.7655        &  0.97048         &  0.9442         &  0.9614 \\
    Cross-Modality \cite{li2015cross} &    -             &  0.7569        &  0.9816          &  0.9527         &  0.9738 \\
    U-net \cite{ronneberger2015u}           &    0.8142        &  0.7537        &  \textbf{0.9820} &  0.9531         &  0.9775 \\
    Deep Model \cite{liskowski2016segmenting}     &    -             &  0.7763        &  0.9768         &  0.9495          &  0.9720 \\
    RU-net \cite{alom2018recurrent}         &    0.8149        &  0.7726        &  0.9820         &  0.9553          &  0.9779 \\
    R2U-net \cite{alom2018recurrent}        &    0.8171        &  0.7792        &  0.9813         &  0.9556          &  0.9782 \\
    \midrule
    BCDU-net (d=1) \cite{azad2019bi}   &    0.8222       & \textbf{0.8012} &  0.9784         &  0.9559          &  0.9788 \\
    BCDU-net (d=3) \cite{azad2019bi}  & \textbf{0.8224}  &  0.8007        &  0.9786         & \textbf{0.9560}  &  0.9789 \\
    \midrule
    InceptNet (d=1) &   0.8137        &  0.7727        &  0.9816         &  0.9550          &  0.9766 \\
    InceptNet (d=3) &   0.8205        &  0.7987        &  0.9783         &  0.9555 &  \textbf{0.9789} \\
    \bottomrule
  \end{tabular}
  \label{tab:table_3}
\end{table}

\subsection{Skin Lesion Segmentation}
\begin{figure}[tb]
    \centering
    \includegraphics[width=13cm, height=3.5cm]{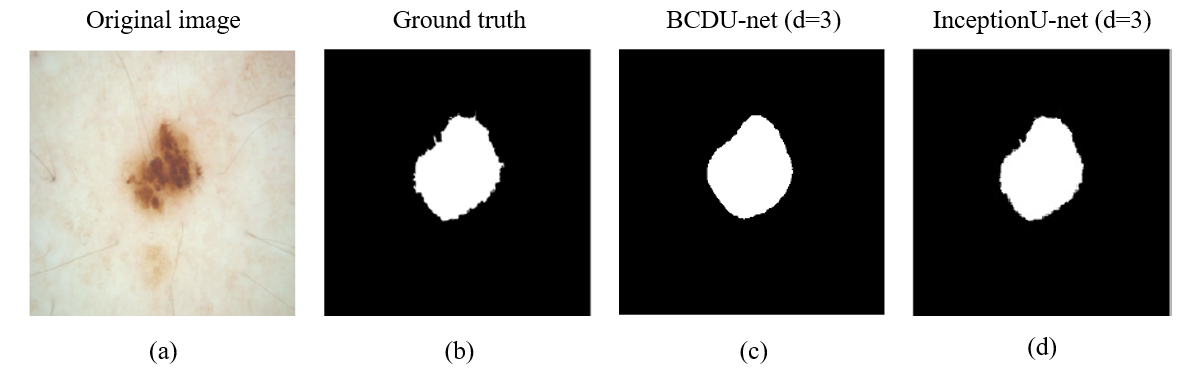}
    \caption{a)original colored image, b) corresponding ground truth mask, c) predicted mask by BCDU-net (d=3), d) predicted mask by InceptNet (d=3) on the ISIC dataset}
    \label{Figure_7} 
\end{figure}

    As for the DRIVE dataset, an example of the output of the proposed network and BCDU-net on the ISIC dataset is illustrated in Figure \ref{Figure_7}. Also, the quantitative results are listed in Table \ref{tab:table_5} obtained by different methods. Again, the InceptNet shows improvement over the original BCDU-net for most of the evaluation metrics. Also, the network with three dense connections outperforms the other one. The convergence plot of the model on the ISIC dataset is shown in Figures \ref{Figure_10}b and \ref{Figure_11}b. According to this plot, the model converges properly. The validation accuracy over the training process is variable. The reason behind this fact is that the validation set contains some images totally different from the ones in the training set, therefore, during the first learning iterations the model has some problems in segmenting those images. In the end, the ROC curve for this dataset is shown in Figures \ref{Figure_12}b and \ref{Figure_13}b.\par

\begin{table}
 \caption{Performance of the different segmentation methods on the ISIC dataset.}
  \centering
  \begin{tabular}{llllll}
    \toprule
    Method                 & F1-score       & Sensitivity   & Specificity     & Accuracy       & JS\\
    \midrule
    U-net \cite{ronneberger2015u}              &    0.647       &  0.708        &  0.964         &  0.890         &  0.549 \\
    Attention U-net \cite{oktay2018attention}   &    0.665       &  0.717        &  0.967         &  0.897          &  0.566 \\
    R2U-net \cite{alom2018recurrent}           &    0.679       &  0.792        &  0.928         &  0.880          &  0.581 \\
    Attention R2U-net \cite{alom2018recurrent} &    0.691       &  0.726        &  0.971         &  0.904          &  0.592 \\
    \midrule
    BCDU-net (d=1) \cite{azad2019bi}      &    0.847       &  0.783        &  0.980         &  0.936          &  0.936 \\
    BCDU-net (d=3) \cite{azad2019bi}     &    0.851       &  0.785        &  0.982         &  0.937          &  0.937 \\
    \midrule
    InceptNet (d=1)   &   0.875        &  0.832        &  0.975         &  0.936          &  0.937 \\
    InceptNet (d=3) & \textbf{0.902} & \textbf{0.857} & \textbf{0.985} & \textbf{0.951} & \textbf{0.951} \\
    \bottomrule
  \end{tabular}
  \label{tab:table_4}
\end{table}


    As for the DRIVE dataset, the proposed architecture could better capture the subtle structures in the image as shown in Figure \ref{Figure_7}. As can be seen in this figure, the proposed method was able to capture the variations in the regions of interest. However, the BCDU-net architecture provided a smoother output (figure \ref{Figure_7}c).\par

\subsection{Lung Nodule Segmentation}
\begin{figure}[tb]
    \centering
    \includegraphics[width=13cm, height=3.5cm]{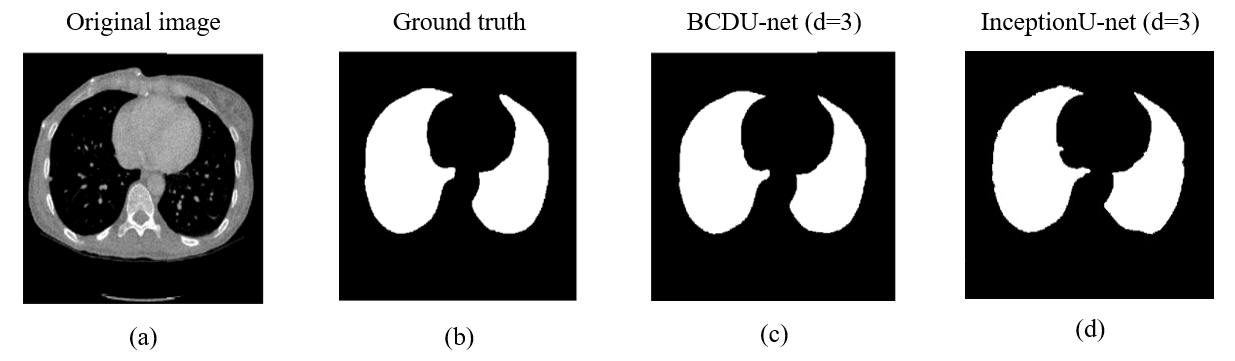}
    \caption{a) Original input image, b) corresponding ground truth mask, c) predicted mask by BCDU-net (d=3), d) predicted mask by InceptNet (d=3) on the Lung dataset}
    \label{Figure_8}
\end{figure}

An example of the qualitative output of both networks (BCDU-net and InceptNet) over the lung dataset is shown in Figure \ref{Figure_8}. Moreover, like the previous datasets, the quantitative results in comparison with other methods are presented in Table \ref{tab:table_6}. For this dataset, the proposed architecture could not improve the results of the BCDU-net structure. This is due to the uniform structure of the region of interest in this dataset. The convergence curve and the ROC curves are depicted in Figures \ref{Figure_10}c, \ref{Figure_11}c, and Figures \ref{Figure_12}c, and \ref{Figure_13}c respectively. \par

\begin{table}
 \caption{Performance of the different segmentation methods on the Lung dataset.}
  \centering
  \begin{tabular}{lllllll}
    \toprule
    Method               & F1-score      & Sensitivity   & Specificity     & Accuracy   & AUC     & JS\\
    \midrule
    U-net \cite{ronneberger2015u}            &    0.9658     &  0.9696       &  0.9872        &  0.9872     & 0.9784  &  0.9858 \\
    RU-net \cite{alom2018recurrent}          &    0.9638     &  0.9734       &  0.9866        &  0.9836     & 0.9800  &  0.9836 \\
    R2U-net \cite{alom2018recurrent}         &    0.9832     &  0.9944       &  0.9832        &  0.9918     & 0.9889  &  0.9918 \\
    \midrule
    BCDU-net (d=1) \cite{azad2019bi}    &    0.9889     &  0.9901       &  0.9979        &  0.9967     & 0.9940 &  0.9967 \\
    BCDU-net (d=3) \cite{azad2019bi} &\textbf{0.9904} & \textbf{0.9910} & \textbf{0.9982} & \textbf{0.9972} & \textbf{0.9946} & \textbf{0.9972} \\
    \midrule
    InceptNet (d=1) &   0.9843      &  0.9874       &  0.9960        &  0.9940     & 0.9915 &  0.9940 \\
    InceptNet (d=3) &   0.9840      &  0.9816       &  0.9962        &  0.9945     & 0.9919 &  0.9945 \\
    \bottomrule
  \end{tabular}
  \label{tab:table_5}
\end{table}


According to Figure \ref{Figure_8}9, the proposed method shows worse performance than the BCDU-net architecture as the proposed method captures unnecessary and very subtle variations in the regions of interest. Since the ground truth mask has a smooth structure, the original BCDU-net is a better candidate for estimating the mask.\par

\subsection{Breast Cancer Cell Segmentation}
\begin{figure}[tb]
    \centering
    \includegraphics[width=13cm, height=3.5cm]{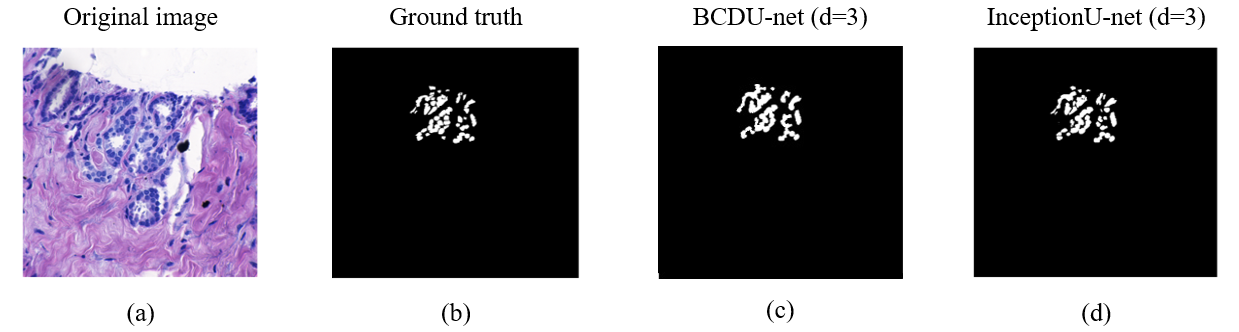}
    \caption{a) Original input image, b) corresponding ground truth mask, c) predicted mask by BCDU-net (d=3), d) predicted mask by InceptNet (d=3) on the Breast Cancer Cell dataset}
    \label{Figure_9}
\end{figure}

\begin{figure}[tb]
         \centering
         \includegraphics[width=10.5cm, height=6cm]{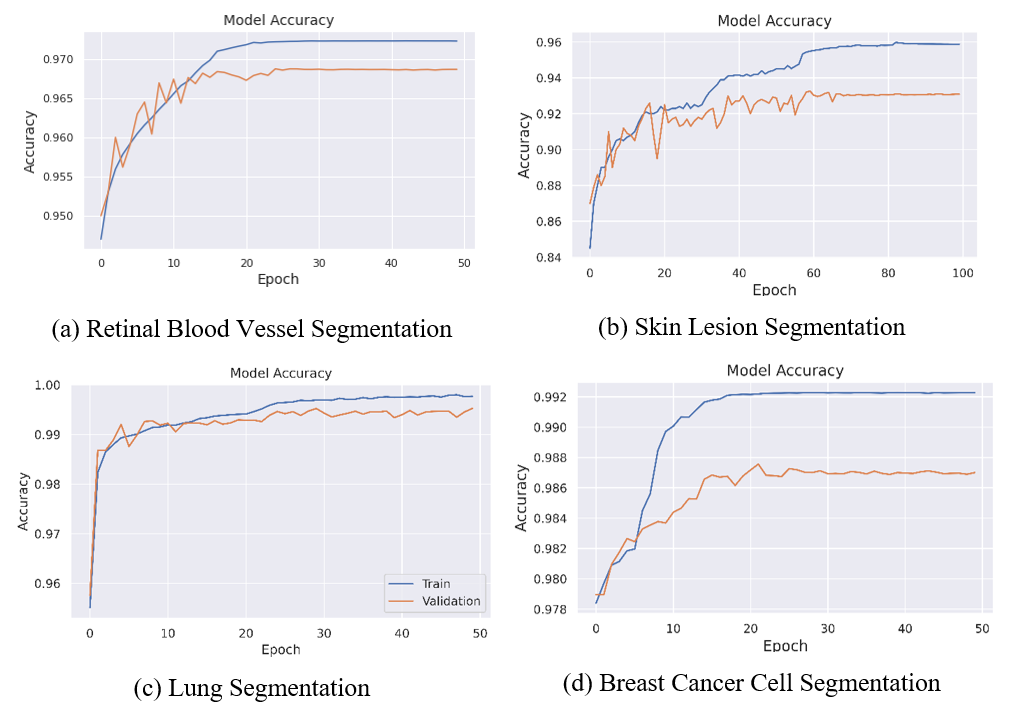}
         \caption{Convergence curves of the InceptNet architecture without dense connections for each dataset}
         \label{Figure_10}

         \centering
         \includegraphics[width=10.5cm, height=6cm]{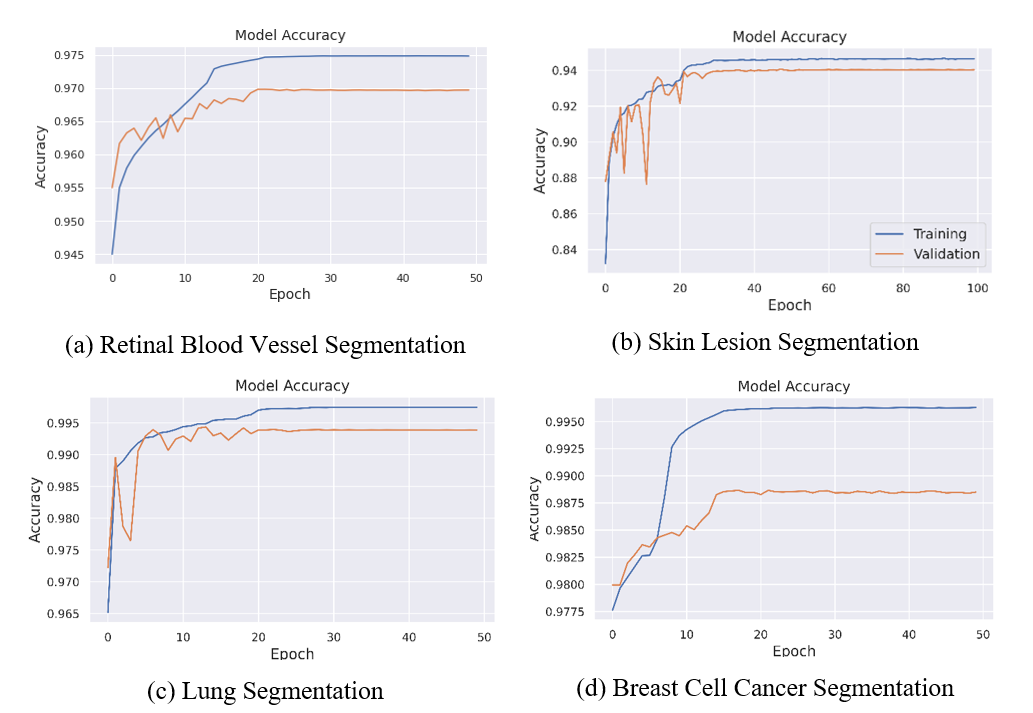}
         \caption{Convergence curves of the InceptNet architecture with dense connections for each dataset}
         \label{Figure_11} 
\end{figure}

\begin{figure}[tb]
         \centering
         \includegraphics[width=10.5cm, height=6cm]{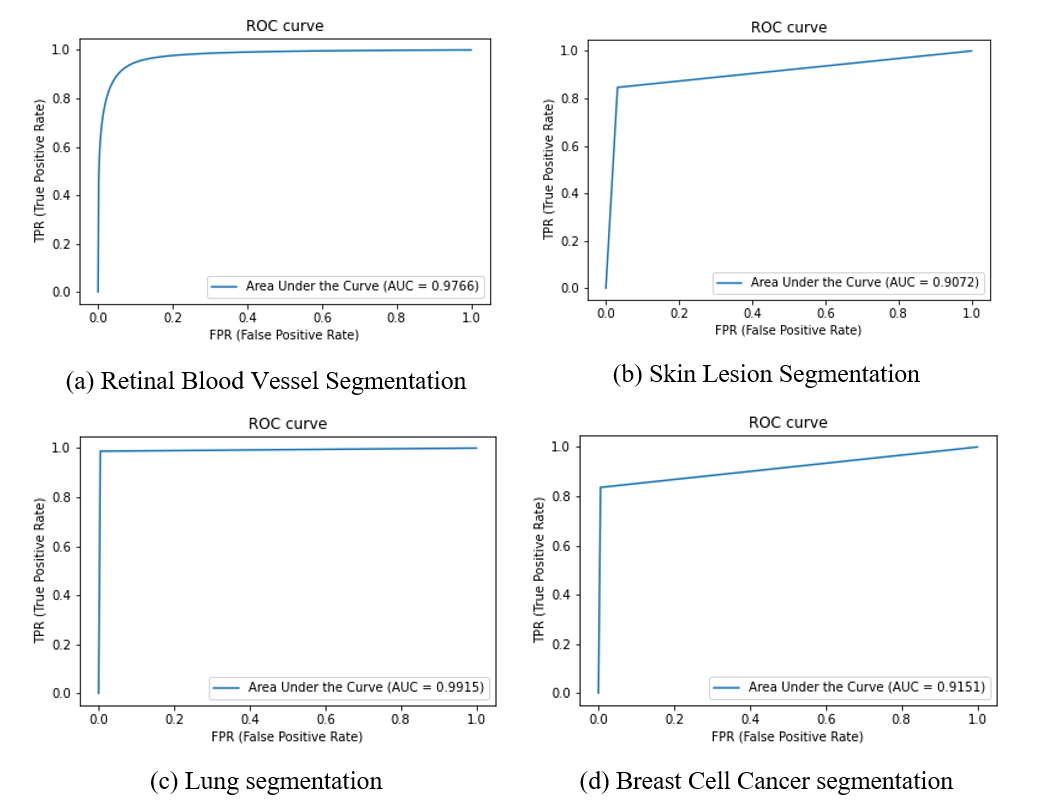}
         \caption{ROC curves of the InceptNet architecture without dense connections for each dataset.}
         \label{Figure_12}
\end{figure}

\begin{figure}[tb]
         \centering
         \includegraphics[width=10.5cm, height=6cm]{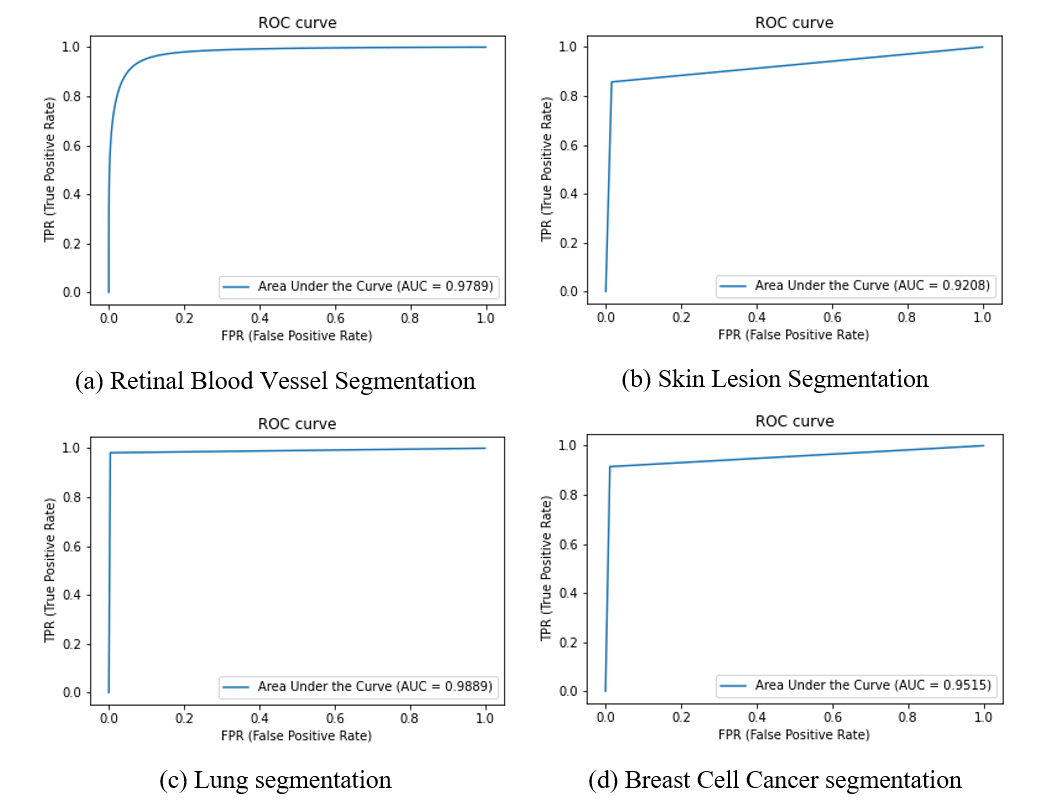}
         \caption{ROC curves of the InceptNet architecture with dense connections for each dataset.}
         \label{Figure_13}
\end{figure}

The final task is the segmentation of breast cancer cells. Figure \ref{Figure_9} shows the qualitative output of the BCDU-net and InceptNet for this dataset. As the BCDU-net with d = 3 had better performance than d = 1, the results obtained by d = 3 are illustrated for both networks (BCDU-net and InceptNet) in Figure \ref{Figure_9}. Table \ref{tab:table_7} lists the quantitative results obtained from different models. Results show that the proposed architecture achieves better results. Also, the network with dense connections performs better in comparison to the network without dense connections. The training and validation accuracy and ROC curves are shown in Figures \ref{Figure_10}d, \ref{Figure_11}d, and Figures \ref{Figure_12}d, and \ref{Figure_13}d, respectively.\par

\begin{table}
 \caption{Performance of the different segmentation methods on the Breast Cell Cancer dataset.}
  \centering
  \begin{tabular}{lllllll}
    \toprule
    Method               & F1-score    & Sensitivity   & Specificity     & Accuracy   & AUC     & JS\\
    \midrule
    U-net \cite{ronneberger2015u}            &   0.6955  & \textbf{0.9331} &  0.9868        &  0.9881     & 0.9444  &  0.9858 \\
    \midrule
    BCDU-net (d=1) \cite{azad2019bi}    &  0.7489     &  0.8542       &  0.9906        &  0.9765     & 0.9318 &  0.9765 \\
    BCDU-net (d=3) \cite{azad2019bi}   &  0.7515     &  0.8982       &  0.9913        &  0.9897     & 0.9448 &  0.9897 \\
    \midrule
    InceptNet (d=1) & \textbf{0.7749} & 0.8956    &  0.9924        &  0.9916     & 0.9151 &  0.9916 \\
    InceptNet (d=3) &   0.7601    &  0.9146       & \textbf{0.9930} & \textbf{0.9945} & \textbf{0.9515} & \textbf{0.9945}  \\
    \bottomrule
  \end{tabular}
  \label{tab:table_6}
\end{table}
    
As Figure \ref{Figure_9} shows, the original BCDU-net completely fails at capturing the small-scale regions of interest and has a smooth and approximate estimation of the ground truth mask. However, the proposed approach could effectively mitigate this issue by the use of Inception modules with convolution layers of different kernel sizes in parallel. \par
    
\begin{figure}[tb]
    \centering
    \includegraphics[width=12cm,height=8cm]{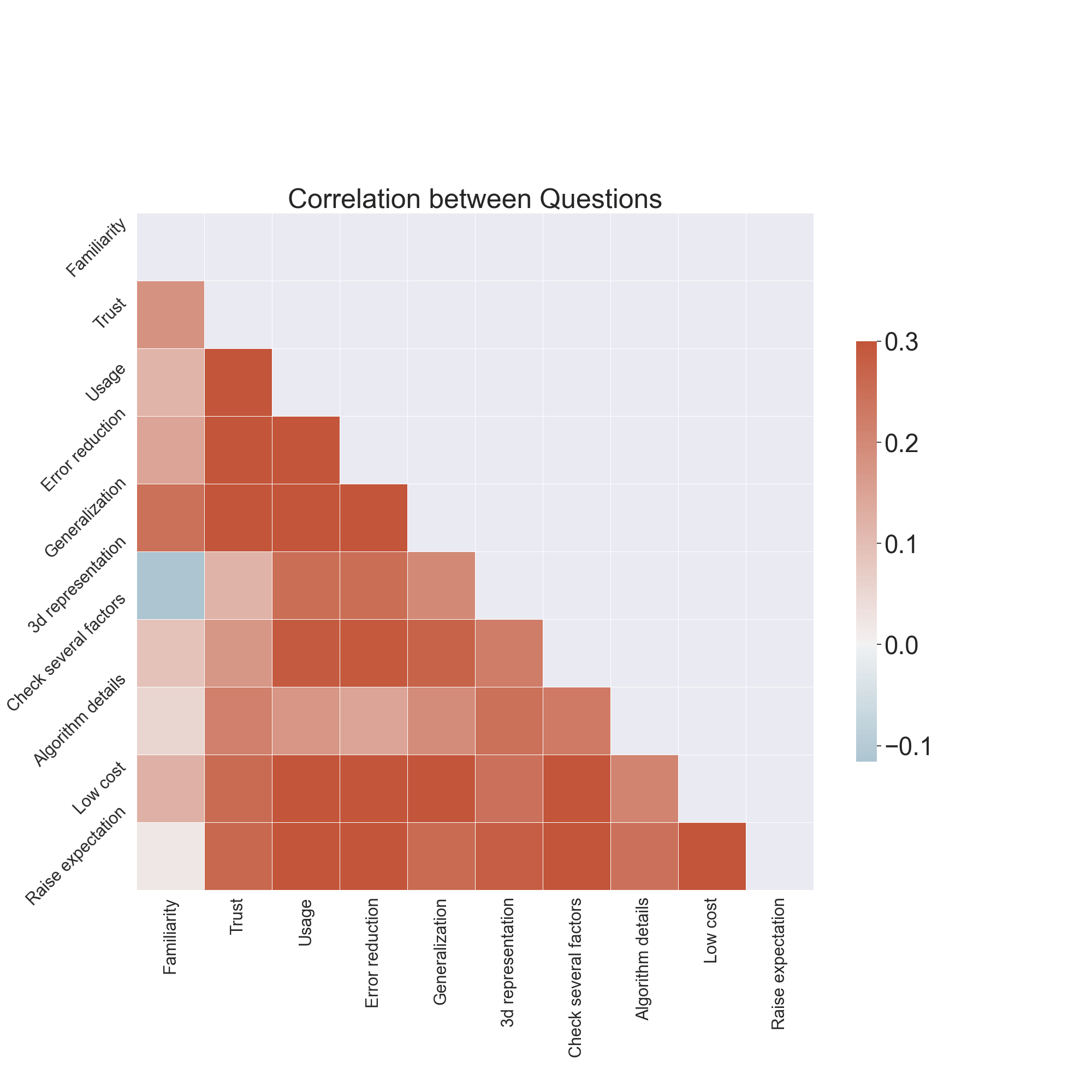}
    \caption{Correlation Matrix between users' answers}
    \label{fig:corr}
\end{figure}

\section{Discussion}
\begin{figure} [tb]
\begin{center}
\includegraphics[width=17cm, height=4cm]{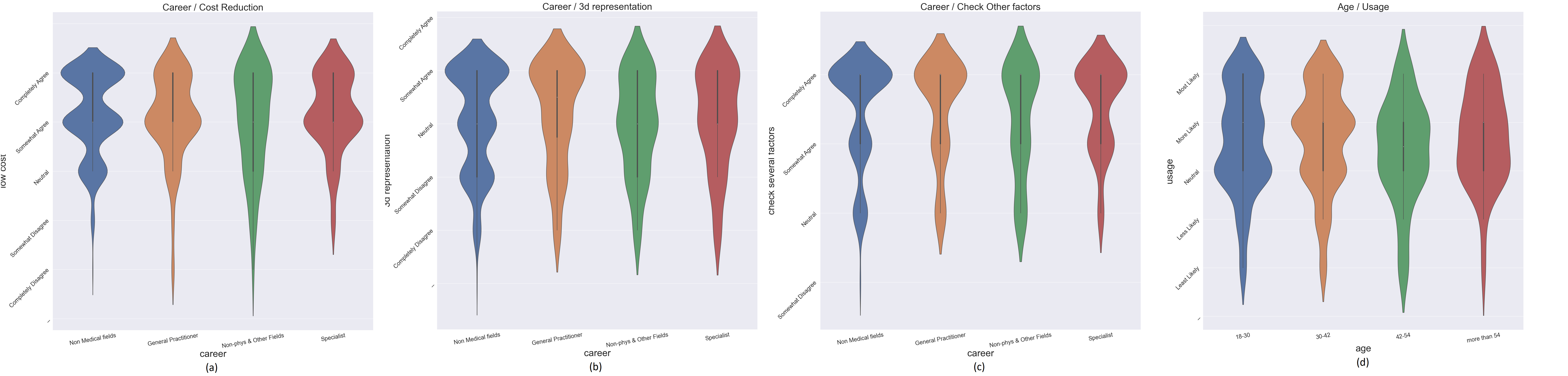}
\end{center}
\caption{a) shows the correlation between career and cost reduction. b) the correlation between career and 3D representation. c) correlation between career and check other factors. d) correlation between age and usage
}
\label{Figure_10.1}
\end{figure}

Analysis of the 538 responses in the collected data identified several key insights around user perceptions of Incept-Net (the list of questions is presented in Appendix 1), as summarized below:  The correlation matrix between opinion answers is depicted in Figure \ref{Figure_10.1}.
\begin{itemize}
    \item Familiarity and Trust are strongly correlated with overall opinions. Respondents with higher familiarity and trust had much more positive views of benefits like error reduction. This aligns with technology acceptance models showing perceived usefulness and ease of use drive adoption \cite{venkatesh2016unified}.
    \item Younger users are more open to using Incept-Net and trusting it. The 18-30 group had a higher willingness to use and trust, while older groups were more hesitant. This generational difference in tech adoption is seen across many studies \cite{mihailidis2022older}.
    \item Non-medical users showed more caution around capabilities like checking multiple factors. They tended to be more neutral compared to medical users. Customizing features to address domain needs may broaden appeal. 
    \end{itemize}

The above points can be observed in Figure \ref{Figure_10.1} which is depicted in Violin Plot form. It shows the distribution of data based on categories of the answers between two features. To build on these findings, some areas for further study include: \par
   
Conducting focus groups with different demographics to delve into adoption barriers like trust. Qualitative data can uncover subtle user concerns missed in closed-ended surveys. Testing interfaces tailored to different use cases and age groups. Observational studies with prototypes can reveal usability issues before launch. Partnering with key industries to pilot Incept-Net for their needs. Field studies produce actionable feedback on integrating AI into real-world workflows. \par

\subsection{Method}
According to Tables \ref{tab:table_4} to \ref{tab:table_7}, the proposed methods show promising results in the segmentation of medical images. On the DRIVE dataset, the proposed methods show much better performance in most of the evaluation metrics over the U-net and BCDU-net architectures. Although the U-net model has a higher specificity, there is a big difference between the specificity and sensitivity values which is a negative point. Also, Figure \ref{Figure_6} shows that the proposed method can better deal with the scale variations in the regions of interest thanks to the Inception modules. About the ISIC dataset, the proposed method has achieved better results in all of the evaluation metrics in comparison with other segmentation methods. Like the DRIVE dataset, the proposed method could capture the tiny structures that the BCDU-net was unable to capture. On the other hand, the best performance on the Lung dataset was for the original BCDU-net architecture. However, InceptNet with dense connections has shown better performance than InceptNet without dense connections. Since this dataset has a smoother structure in the regions of interest, Inception modules capture some tiny structures that are not a part of these regions and consequently increase the error of detection. For the last dataset (Breast Cancer Cell), the proposed network achieves far better results regardless of dense connections. Although the U-net has the highest sensitivity value, the value of the F1-score is much lower compared to other methods which is a big disadvantage for this method. As can be seen in Figure \ref{Figure_9}, this dataset consists of images with lots of small structures that can deceive the network. As a result, a high number of Inception modules may label non-cancerous pixels as positive pixels. On the other hand, if there is no Inception module, there is a possibility of missing some positive pixels (cancerous cells). As the results show, the proposed method outperforms the BCDU-net structure as it can better capture the small structures because of the Inception modules in the convolutional blocks. However, the InceptNet with dense connections (d=3) has a worse performance in terms of accuracy and F1-score (table \ref{tab:table_7}) which is because of a higher number of trainable parameters that can lead to overfitting (figure \ref{Figure_11}d). Consequently, if the dataset has a large number of tiny structures, it is better to use the network without dense connections in the last layer of the encoding path.\par
    To have a better comparison between the performance of the models on the datasets, the summary of the accuracy and F1-score is presented in Table \ref{tab:table_7}. It can be seen that the results of BCDU-net are improved on the DRIVE, ISIC 2018, and Breast Cell Cancer datasets by replacing Inception blocks with a series of two convolutional layers. Also, the results of the proposed method on the Lung dataset are close to the results of the original BCDU-net architecture despite the fact that the number of learning parameters of the proposed model is fewer than the number of parameters of the BCDU-net model.\par

\begin{table}
 \caption{Comparison of the performance of the BCDU-net and the proposed method}
  \centering
  \begin{tabular}{lllllllll}
    \toprule
    \multicolumn{9}{c}{   -------------------------------DRIVE---------------------ISIC-----------------------Lung-------------------Breast Cancer}                \\
    \cmidrule(r){1-9}
    Method & F1-score & Accuracy & F1-score & Accuracy & F1-score & Accuracy & F1-score & Accuracy \\
    \midrule 
    U-net \cite{ronneberger2015u}  & 0.8142 & 0.9531 & 0.6470 & 0.8900 & 0.9658 & 0.9872 & 0.6955 & 0.9881 \\
    BCDU-net (d=1) \cite{azad2019bi} & 0.8222 & 0.9559 & 0.8470 & 0.9360 & 0.9889 & 0.9967 & 0.7489 & 0.9765 \\
    BCDU-net (d=3) \cite{azad2019bi} & \textbf{0.8224} & \textbf{0.9560} & 0.8506 & 0.9374 & \textbf{0.9904} & \textbf{0.9972} & 0.7515 & 0.9897 \\
    InceptNet (d=1) & 0.8137 & 0.9550 & 0.8749 & 0.9358 & 0.9843 & 0.9940 & \textbf{0.7749} & 0.9916 \\
    InceptNet (d=3) & 0.8205 & 0.9555 & \textbf{0.9020} & \textbf{0.9510} & 0.9840 & 0.9945 & 0.7601 & \textbf{0.9945} \\ 
    \bottomrule
  \end{tabular}
  \label{tab:table_7}
\end{table}

    As mentioned in Section 4, using convolutional layers with different kernel sizes in parallel could capture the regions of interest with different scales. Consequently, the proposed method showed a better result compared to the BCDU-net architecture on the datasets with regions of interest of various scales, such as the DRIVE, ISCI 2018, and Breast Cancer Cell datasets. However, according to Table \ref{tab:table_7}, the original BCDU-net showed the best performance on the Lung dataset. This is because of the fact that the variation in the scale of the regions of interest in this dataset was lower compared to the other datasets. As a result, the Inception block with different kernel sizes does not make an improvement on the results of the segmentation task on this dataset. Also, using dense connections can improve the results if the dataset under study has small-scaled structures but not a lot.\par

\section{Conclusion}

In this study, InceptNet was proposed for segmenting medical images and detecting disease at an early stage. In particular, if required clinical information is available, the proposed model can detect the diseases even in the early stage and enhance the accuracy of treating the disease. To obtain information on how the model performs, this model was tested on four benchmark datasets. Results showed that incorporating Inception blocks with different kernel sizes in parallel improved the segmentation on the datasets where there were variations in the regions of interest in the ground truth. Also, in most cases, dense connections at the end of the network improve precision and help information flow over the layers. If the distribution of small-scale structures is dense, using dense connections in the last layer of the encoding path can lead to overfitting and worsening the results. Overall, this result highlighted the potential of the model as a practical tool for disease detection and segmentation of medical images. \par
In summary, the questionnaire provides a foundation to guide ongoing HCI research around Incept-Net. Combining survey data with further qualitative, observational, and field studies provided a comprehensive view of users' needs to successfully develop and deploy the application.

\textbf{Funding}
    The authors declare that no funding was received to assist with the preparation of this manuscript.

\textbf{Conflict of Interest}
    The authors declare that they have no known competing financial interests or personal relationships that could have appeared to influence the work reported in this paper.

\textbf{Data availability}
    The datasets analyzed during the current study are publicly available and their links have been provided in the footnotes.


\bibliographystyle{unsrt}  
\bibliography{references}  
\newpage

\appendix
\section{Appendices}
\subsection{}

The designed questionnaire comprised 13 inquiries that encompassed both individual information and user perspectives pertaining to their interaction with a trial version of InceptNet in the shape of an application, our novel methodology. It is a fundamental tenet that every system is tailored to cater to a specific community, and thus, we endeavored to gain a comprehensive understanding of this target audience to improve our design process. \par

The primary objective of this questionnaire was to discern avenues for enhancing human perception and user experience during the utilization of our proposed method. Moreover, we sought to investigate the efficacy of presenting segmented images to 538 operators hailing from diverse academic disciplines. The questionnaire probed users' sentiments, including aspects such as trust, to glean valuable insights into their subjective impressions. The questions come as follows with their possible answers to select.\par
\begin{itemize}
    \item Gender Identification
    \item Age 
    \item Career field 
    \item Level of familiarity with the AI assistant of the medical image analyst
    \item In diagnosing diseases from medical images, I will trust AI assistants
    \item The proposed Inceptnet AI application is able to accurately diagnose 95-99\% of diseases identified from medical images Would you like to use this smart system instead of classical disease diagnosis methods?
    \item I would probably use a medical image analysis AI assistant to reduce the chance of error in diagnosis
    \item AI assistant usage will be generalized for disease diagnosis among doctors will be obvious
    \item For increasing the accuracy of disease diagnosis, 3D representation of disease diagnosis in images will be more effective than the result of 2D images
    \item In order to increase the accuracy of disease diagnosis, it is better for the Inceptnet AI assistant to check several other influencing factors in the patient at the same time
    \item In the not-so-distant future, the diagnosis of the disease will be made with 100\% accuracy and without errors by the Inceptnet  AI assistant
    \item The correct diagnosis of the disease with Inceptnet AI assistant help in the early stages is an effective help considering the low cost
    \item The result of disease diagnosis in medical images with the help of Inceptnet AI assistant helps me to diagnose the disease and its progress stages
\end{itemize}

\end{document}